\documentclass[11pt,letterpaper]{article}

\usepackage{jheppub}
\usepackage{amsmath}
\usepackage{amsfonts}
\usepackage{amssymb}
\usepackage{graphicx}
\usepackage{epsfig}

\def\ba{\begin{eqnarray}}
\def\ea{\end{eqnarray}}
\def\bea{\begin{eqnarray}}
\def\eea{\end{eqnarray}}
\def\be{\begin{equation}}
\def\ee{\end{equation}}

\def\({\left(}
\def\){\right)}
\def\[{\left[}
\def\]{\right]}

\title{\center{What is the shape of the initial state?}\footnotemark[1]\footnotetext[1]{Essay awarded second prize in the Gravity Research Foundation 2013 essay competition}}

\author[a]{Nishant Agarwal,}
\author[b]{R.~Holman,}
\author[c]{and Andrew~J.~Tolley}

\affiliation[a]{McWilliams Center for Cosmology, Department of Physics, Carnegie Mellon University, Pittsburgh, PA 15213, USA}
\affiliation[b]{Department of Physics, Carnegie Mellon University, Pittsburgh, PA 15213, USA}
\affiliation[c]{Department of Physics, Case Western Reserve University, 10900 Euclid Ave, Cleveland, OH 44106, USA}

\emailAdd{nishanta@andrew.cmu.edu}
\emailAdd{rh4a@andrew.cmu.edu}
\emailAdd{andrew.j.tolley@case.edu}

\abstract {We argue that a plausible operational definition for an initial state of the Universe is the initial quantum state of the curvature perturbations generated during inflation. We provide a parameterization of this state and generalize the standard in-in formalism to incorporate the structures in this state into the computation of correlators of the perturbations. Measurements of these correlators using both the CMB as well as large scale structure probe different structures in the initial state, as they give rise to bi- and tri-spectra peaked on different shapes of triangles and quadrilaterals in momentum space. In essence, the shapes implied by the correlators feed directly into information about the shape of the initial state and what physics could have preceded inflation to set this state up.}

\begin{document}

\maketitle

\newpage


How can we access information about the initial state of the Universe? First of all it would be useful to have a definition of this state which, all parties can agree, represents something that makes sense as an initial state. Part and parcel of arriving at such a definition would be the requirement that this state can be probed experimentally.

Our main signpost in the search for an initial state is inflation \cite{Guth:1980zm,Linde:1981mu,Albrecht:1982wi}, which has become the dominant paradigm to explain the large-scale homogeneity of the Universe and generate the initial perturbations that seeded cosmic structure. It is also becoming increasingly possible to probe the physics of inflation using excellent cosmological observations from the cosmic microwave background (CMB) and large scale structure (LSS).

In an ideal world, we would have access to a full theory of quantum gravity at both weak and strong coupling and be able to input test initial quantum states for the relevant variables in such a theory, calculate observables emerging from these initial conditions, and then compare those results to observations. Unfortunately, we are far from living in such a world and must rely on proxies, such as quantum field theory defined in a curved spacetime. This technology \cite{BirrellDavies,ParkerToms} has been used to predict the evaporation of black holes, as well as to calculate the quantum fluctuations of fields in an inflationary cosmology \cite{Mukhanov:1981xt,Starobinsky:1982ee,Hawking:1982cz,Guth:1982ec,Linde:1982uu,Bardeen:1983qw}; it is this latter application that will be of interest to us below.

This leads us to our definition of the initial state of the Universe: {\it it is the quantum state of the curvature perturbations generated during inflation}. Now, it is easy to argue that there is surely a pre-inflationary era and any initial state must refer to this era. But inflation is the great eraser; given a sufficient number of e-folds, pre-inflationary effects can be suppressed to unobservable levels. Thus, from an operational point of view, if inflation happened at all, an event that seems more and more likely with the release of every new cosmological data set, our ability to infer the state of the early universe will be limited to what we can say about the initial state of the curvature perturbations. However, information about this state can then be used to, at the very least, constrain pre-inflationary physics.

Our goal in this essay is to discuss how an effective theory of initial states, similar in philosophy to effective field theories used to understand possible physics beyond the standard model, can be developed. This, used in conjunction with the extant effective theory of inflation \cite{Cheung:2007st},  would allow us to calculate the effect of structure beyond that of the free vacuum in the inflaton initial state.

\section{An effective field theory of the initial state}

One potential barrier in using cosmological data to pin down the initial state has to do with the gamut of observationally consistent inflationary models. Even within the realm of single-field inflation, many different models are consistent with the current data. How then can we decouple model-specific effects from those due to the initial state? What is needed is a unifying description for different models of inflation --- an effective field theory approach is therefore ideally suited.

An effective field theory of inflation was developed recently by identifying fluctuations of the inflaton with the Goldstone mode of spontaneously broken time translations \cite{Cheung:2007st}. With this technology in hand, we can develop a method for parameterizing the initial density matrix and then incorporating it into an effective action. Once this is done, we can then use the in-in formalism \cite{Schwinger:1960qe,Keldysh:1964ud,Bakshi:1962dv,Bakshi:1963bn} to calculate the effects of non-trivial initial states on the correlation functions of metric perturbations \cite{Agarwal:2012mq}.

\begin{figure}[!h]
  \begin{center}
    \includegraphics[width=5.0in,angle=0]{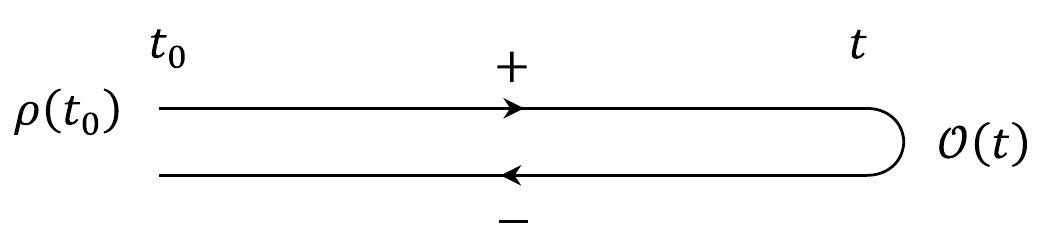}
    \caption{The closed time path contour that forms the basis of the in-in formalism.}
    \label{fig1}
  \end{center}
\end{figure}

The in-in or closed time path formalism is useful to evaluate time-dependent expectation values of operators, as opposed to S-matrix elements connecting in and out states. For instance, the expectation value of an operator  ${\cal O}(t)$ at a time $t$ in the Schr\"odinger picture,
\bea
	\langle {\cal O} \rangle (t) & = & {\rm Tr} \( \rho(t) {\cal O}(t) \) \\
	& = & {\rm Tr} \( U^{\dagger}(t, t_{0}) \ {\cal O}(t) \ U(t, t_{0}) \ \rho(t_{0}) \),
\eea
can be read as follows: start with the initial density matrix $\rho(t_{0})$, evolve it using the operator $U(t,t_{0})$ to a time $t$ at which point the operator ${\cal O}(t)$ is inserted, and finally evolve back to $t_{0}$ (fig. \ref{fig1}). We can explicitly evaluate the trace in the field representation of the field operator $\Phi \( \vec{x},t \)$ and write $\langle {\cal O} \rangle (t)$ as a path integral. Doing so yields a generating functional ${\cal Z} \[ J^{+}, J^{-}; t_{0} \]$ of the form
\bea
	{\cal Z} \[ J^{+}, J^{-}; t_{0} \] & = & \int {\cal D}\Phi^{+} \ {\cal D}\Phi^{-} \exp \Big( i \( S\[ \Phi^{+}, J^{+} \] - S\[ \Phi^{-}, J^{-} \] \) \Big) \ \rho\(\Phi^{+}, \Phi^{-}; t_{0} \), \quad
\label{eq:genfcnl}
\eea
where the plus and minus fields are evaluated on the forward and backward parts of the time contour respectively and $S \[ \Phi^{\pm}, J^{\pm} \]$ is the action of the system, having added sources $J^{\pm}$ to the appropriate pieces.

As it stands, the generating functional in eq. (\ref{eq:genfcnl}) is not yet written in a way where we can calculate the correlation functions of interest. To do so, let us parameterize the initial state as $\rho\(\Phi^{+}, \Phi^{-}; t_{0} \) = N {\rm exp} \( i {\cal S} \[ \Phi^{+}, \Phi^{-}; t_{0} \] \)$, $N$ being the normalization chosen such that ${\rm Tr} \( \rho(t_{0}) \) = 1$, and ${\cal S}$ an action that inherits properties from physics above the cutoff. Without loss of generality we can choose ${\cal S}$ to be quadratic at lowest order, with higher order terms denoting non-Gaussianities in the initial state. We can now combine the kinetic piece of the effective action with the Gaussian part of the initial state (denoting the net operator with ${\cal O}_{k}(t,t')$) and use the saddle point method to solve for the Green's function, ${\cal G}_{k}(t, t^{\prime})$, defined via
\bea
	\int {\rm d} \tau \ {\cal O}_{k}(t,\tau) {\cal G}_{k}(\tau, t^{\prime}) & = & -i \delta(t- t^{\prime}) {\mathbb I}_{2},
\label{eq:greens}
\eea
where ${\mathbb I}_{2} \equiv {\rm diag}(1,1)$. Note that eq. (\ref{eq:greens}) is really a collection of four equations, one for each of the $++, \ +-, \ -+$, and $--$ components of the Green's function.

The initial state is typically chosen to be of the form of the Bunch--Davies vacuum \cite{Bunch:1978yq} with $\rho(t_{0}) = 1$. In \cite{Holman:2007na} two of us considered the possibility of a general pure initial state, with ${\rm Tr} \( \rho^{2}(t_{0}) \) = 1$, given by a Bogoliubov transformation of the Bunch--Davies vacuum. In an effective field theory setting, however, one must account for the high energy degrees of freedom that have been integrated out in order to write a low energy effective action. The initial density matrix for our Universe must therefore be interpreted as a {\it reduced} density matrix, obtained by tracing over the multiverse. Since the heavy degrees of freedom will in general be entangled, we expect the initial state of our Universe to be a mixed state, with ${\rm Tr} \( \rho^{2}(t_{0}) \) < 1$ (see e.g. \cite{GottfriedYan}). In \cite{Agarwal:2012mq} (also see \cite{Agullo:2010ws,Agullo:2011xv}) we constructed the Green's function for such general initial states and used it to calculate higher-order correlation functions of metric perturbations. Matching these predictions to CMB and LSS data yields a mechanism to constrain the initial state.

\section{Constraining the initial state with cosmological data}

Given a formalism to compute the Green's function that incorporates generic initial state effects, the connection to observations will come from computing correlation functions of the primordial curvature perturbation $\zeta(t,\vec{x})$ (which is simply related to the Goldstone mode of the effective field theory). While the power spectrum sets the overall scale of inhomogeneities produced during inflation, higher-order correlation functions directly probe the dynamics and interactions of the inflaton. We expect, therefore, that one must look at $N$ point ($N > 2$) functions of metric perturbations to constrain the initial state. Of course as $N$ becomes larger, the effects become smaller and hence more difficult to observe. As such, the existence of Ward identities for conformal symmetries, that relate $N$ to $N-1$ point functions, can prove to be very relevant to our understanding of inflation \cite{Assassi:2012zq,Hinterbichler:2012nm,Mata:2012bx,Creminelli:2012qr,Hinterbichler:2013dpa}. The well-known consistency relation for single field inflation \cite{Maldacena:2002vr,Creminelli:2004yq} between the bispectrum and the power spectrum may be an example of such a symmetry. We find that general initial states can violate the consistency condition, though it is yet to be seen how they affect the full set of Ward identities.

Let us consider the three-point function $\big\langle \zeta_{\vec{k}_{1}} \zeta_{\vec{k}_{2}} \zeta_{\vec{k}_{3}} \big\rangle = \(2\pi\)^{3} \delta^{3}\big(\sum \vec{k}_{i}\big) B_{k_{1},k_{2},k_{3}}$, with $B_{k_{1},k_{2},k_{3}}$ being the bispectrum. In the formalism discussed in the previous section, the three-point function can be calculated using any combination of plus and minus fields. The full bispectrum is computed after including cubic terms from the effective field theory action, and also from the initial state if considering initial non-Gaussianities. One can then perform Wick's contractions to write the bispectrum in the form of integrals over products of Gaussian Green's functions. We refer the interested reader to \cite{Agarwal:2012mq} for details of the calculation and summarize the result here.

We can write the bispectrum in the form $B_{k_{1},k_{2},k_{3}} = {\cal B}_{k_{1},k_{2},k_{3}} P_{k_{1}} P_{k_{3}}$, where we have factored out the power spectrum $P_{k}$ in the modes $k_{1}$ and $k_{3}$. In the squeezed triangle limit $k_{1} \approx k_{2} \gg k_{3}$, ${\cal B}_{k_{1},k_{2},k_{3}} \approx (12/5)f_{\rm NL}^{\rm local}$, where $f_{\rm NL}^{\rm local}$ is the well-studied local non-Gaussianity parameter (usually assumed constant). The consistency relation states that for single-field models of inflation $f_{\rm NL}^{\rm local}$ is of the order of the slow-roll parameters, i.e. ${\cal O}(0.01)$. While this is true for a Bunch--Davies initial state, general initial states show an enhancement in $f_{\rm NL}^{\rm local}$ of the form $k_{1}/k_{3}$ at {\it leading order in slow-roll}. Therefore, not only is the $f_{\rm NL}$ parameter no longer proportional to the slow-roll parameters, it is further enhanced by the ratio of momenta as well. This violation of the consistency relation is testable with CMB and LSS data.

We note that the momentum enhancement in the squeezed limit and the resulting violation of consistency has been observed previously in \cite{Agullo:2010ws,Agullo:2011xv,Ganc:2011dy,Chialva:2011hc}. We report a much stronger violation of the consistency condition since $f_{\rm NL}^{\rm local}$ is no longer slow-roll suppressed.\footnotemark[2]\footnotetext[2]{Recently, \cite{Chen:2013aj} proposed a Lagrangian that violates consistency in a different setting, through an initial non-attractor inflationary phase.} This result should help mitigate the tight constraints on initial state modifications found in \cite{Flauger:2013hra,Aravind:2013lra}. General initial states also lead to an enhancement of the bispectrum in the flattened limit $k_{1} + k_{2} \approx k_{3}$, as has been noted in many works \cite{Chen:2006nt,Holman:2007na,Meerburg:2009ys,Agullo:2010ws,Ganc:2011dy,
Chialva:2011hc,Agarwal:2012mq}.

\section{Conclusions}

The upshot of the above discussion is that the time is ripe, both from the theoretical as well as the observational point of view, to pin down the shape of the initial state. The WMAP and PLANCK CMB data sets coupled with LSS data such as the Sloan Digital Sky Survey (SDSS) will allow us to constrain the shape of the primordial bispectrum to great detail, and thus directly test our ideas. As an example of this, it has already been noted that modified initial states are expected to leave distinct observational signatures in the scale-dependence of the bias of dark matter halos \cite{Agullo:2012cs,Ganc:2012ae}.

We can speculate on what we may find from this analysis. The recent PLANCK release has not seen any evidence for non-Gaussianity of the standard forms (local or equilateral). However, due to the so-called non-factorizable form of the shape that arises from excited initial states, the bounds on the relevant $f_{\rm NL}$ parameter are quite weak at present. There may still be some non-trivial structure beyond the Bunch--Davies vacuum lurking in the data.

If, however, {\em no} evidence for physics beyond the Bunch--Davies state is found, then that tells us a great deal about pre-inflationary physics. Either, the number of e-folds was sufficiently larger than the bare minimum required to solve the horizon and flatness problems to erase this information, which then leads directly to issues in trans-Planckian physics for the modes that are observable today on the CMB sky. Or, (not exclusive) pre-inflationary physics was exceedingly adiabatic so that the initial state suffered no disturbances starting from the free-field vacuum in the deep past. In either case, we can use this information to restrict possible forms of pre-inflationary physics, which in itself would be a great achievement.

\acknowledgments N.~A. was supported by the McWilliams Fellowship of the Bruce and Astrid McWilliams Center for Cosmology. R.~H. was supported in part by the Department of Energy under grant DE-FG03-91-ER40682. A.~J.~T. was supported in part by the Department of Energy under grant DE-FG02-12ER41810.

\bibliography{references}
\bibliographystyle{JHEP}

\end{document}